\documentclass[reprint, aps,prd,twocolumn,showpacs,showkeys, 10pt, longbibliography, nolinenumbers, floatfix]{revtex4-1}

\usepackage{amsmath}
\usepackage{graphicx}
\usepackage{float}
\usepackage{dcolumn}
\usepackage{multirow}
\usepackage{natbib}
\usepackage{bm}
\usepackage{changes}
\usepackage{comment}
\usepackage[colorlinks = true,linkcolor = blue,urlcolor  = blue,citecolor = blue,anchorcolor = blue]{hyperref}
\usepackage{enumerate}

\begin{document}

\title{XTE J1814-338:~A potential hybrid star candidate}

\author{P. Laskos-Patkos}
\email{plaskos@physics.auth.gr}

\author{Ch.C. Moustakidis}
\email{moustaki@auth.gr}

\affiliation{Department of Theoretical Physics, Aristotle University of Thessaloniki, 54124 Thessaloniki, Greece}

\begin{abstract}
A recent analysis on the properties of the XTE J1814-338 pulsar yielded a small radius value around $\sim$ 7 km. Notably, this estimation is significantly lower compared to the currently
inferred values for the radius of neutron stars (as they are derived from both theoretical calculations and corresponding observations). In this paper, we focus on the construction of hybrid equations of state (EOSs) for the possible reconciliation of the exotic XTE J1814-338 properties.~Our analysis indicates that an equation of state involving a sufficiently strong phase transition could potentially lead to the explanation of the size and mass of  such a compact object. By examining the sign of the $dM/dP_c$ slope, we found that stellar configurations compatible to XTE J1814-338 are only stable for a "stiff" low density phase. For a "soft" hadronic EOS selection the stability of the resulting configurations may not be satisfied in terms of the turning point criterion. However, a complementary analysis, for the radial oscillations of the aforementioned configurations, indicated their stability when a slow phase conversion is considered.


\keywords{XTE J1814-338, hadron-quark phase transition, equation of state}
\end{abstract}

\maketitle

\section{Introduction} One of the major unresolved questions regarding the physics of dense nuclear matter is related to the composition of the central region in neutron stars~\cite{Bielich-2020}. In particular, neutron stars could be purely hadronic (composed by nucleons and hyperons), but the extreme conditions that prevail in their interior have allowed physicists to propose more exotic scenarios concerning their structure~\cite{Heiselberg-2000,Weber-2005}. One of these scenarios focuses on the appearance of deconfined quark matter due to a first-order phase transition~\cite{Witten-1984,Heiselberg-2000}.~In that case, a quark matter core is surrounded by a mantle of hadronic matter, and stars with such hybrid structure are named hybrid stars. Interestingly, hybrid stars have shown to fit current gravitational wave constraints and x-ray observations~\cite{Li-2021,Li-2024,Christian-2024,Laskos-Patkos-2024a,Laskos-Patkos-2024b,Kumar-2023}, while several studies have focused on the prediction of signature phenomena that may be associated with their existence~\cite{Sotani-2011,Flores-2014,Prakash-2021,Laskos-Patkos-2023,Lyra-2023,Tan-2022,Christian-2022}.

Recently, the analysis of Kini {\it et al.}~\cite{Kini-2024} provided puzzling values for the mass $(M)$ and radius $(R)$ of the XTE~J1814-338 pulsar.~More precisely, the authors reported that,  at the  $1\sigma$ level, $M=1.21^{+0.05}_{-0.05}$~$M_\odot$ and $R=7.0^{+0.4}_{-0.4}$~km. Up to this moment, different studies have proposed that the presence of dark matter could provide a viable explanation for the existence of the aforementioned ultracompact object.~Specifically, Pitz and Schaffner-Bielich~\cite{Pitz-2024} studied the possibility of XTE J1814-338 being a bosonic star with a nuclear matter core, while Yang {\it et al.}~\cite{Yang-2024} investigated the scenario of a strange star admixed with mirror dark matter. Notably, the idea of ultracompact stars has been studied by Li {\it et al.}~\cite{Li-2023} before the results of Ref.~\cite{Kini-2024}. In particular, the authors showed that hybrid equations of state (EOSs), which are consistent with multi-messenger astronomy (LIGO and NICER measurements), could support stable (in terms of the turning point criterion~\cite{Zel’dovich-1963,Harrison-1965}) hybrid stars with extremely small radii. 

In any case, the existence of this exotic object can, on the one hand, spark numerous speculations about its structure; on the other hand, and perhaps more crucially, it can lead to a revision of our understanding on how such objects form, within an area of the Mass-radius ($M-R$) diagram that remains unexplored and diverges significantly from current theoretical predictions. However, it is important to comment that, as Kini {\it et al.}~\cite{Kini-2024} have stated, the inferred exotic properties could result from systematic errors rooted in our currently limited understanding of burst oscillation mechanisms. Thus, future insight on our comprehension of burst oscillation origins would be of utmost significance.

The purpose of this paper is to examine the conditions under which the XTE~J1814-338 pulsar could be explained as a hybrid star. In addition, we wish to examine the compatibility of the resulting EOSs to current astronomical constraints, including the recent analysis concerning the sub solar mass neutron star in the HESS~J1731-347 remnant~\cite{Doroshenko-2022}. To this end, we have combined two widely employed hadronic models, namely APR~\cite{Akmal-1998} and DD2-GDRF~\cite{Typel-2018}, with the well-known constant speed of sound (CSS) model for quark matter~\cite{Zdunik-2013,Alford-2013,Montana-2019}. We aim to explore how the selected hadronic EOS may impact the reconciliation of the XTE J1814-338 properties, while the CSS parametrization provides a general framework to study the effects of the parameters involved in a first-order phase transition. Finally, it is of utmost importance to examine the stability of the resulting stellar configurations beyond the well-known turning point criterion (which determines the stability based on the slope of the stellar mass with respect to the central pressure)~\cite{Zel’dovich-1963,Harrison-1965}.

\section{ Hybrid equation of state} In the present study, the phase transition is described via Maxwell construction~\cite{Bielich-2020,Heiselberg-2000}. Notably, this approach is the favored one in the case of large surface tension between the hadronic and the quark phase~\cite{Mariani-2017}. In this scenario the phase transition is abrupt, resulting in a density discontinuity. Specifically, the energy density reads~\cite{Blaschke-2013,Alvarez-Castillo,Montana-2019,Alford-2013}
\begin{equation}
  \mathcal{E}(P) = \begin{cases} 
      \mathcal{E}_{\rm HADRON}(P), & P\leq P_{\rm tr} \\
      \mathcal{E}(P_{{\rm tr}}) + \Delta \mathcal{E} + c_s^{-2}(P-P_{{\rm tr}}), & P > P_{{\rm tr}}
   \end{cases}
   \label{1}
\end{equation}
 where $P$ stands for the pressure and $c_s$ is the speed of sound divided by the speed of light ($c_s^2=dP/d\mathcal{E}$). Furthermore, $P_{\rm tr}$ and $\Delta\mathcal{E}$ denote the transition pressure and the energy density jump, respectively. The aforementioned approach is known as the CSS model. It is important to comment that the first line of Eq.~(\ref{1}) refers to the hadronic EOS, while the second one refers to the quark phase. Specifically, the second line of Eq.~(\ref{1}) can be though of as a first-order Taylor expansion of the energy density around the transition pressure. Even though such a treatment lacks a detailed theoretical foundation, it is widely used~\cite{Alford-2014b,Christian-2019,Christian-2021,Christian-2022,Han-2019a,Li-2021,Sharifi-2021,Paschalidis-2018,Alford-2017,Deloudis-2021,Han-2020} as it mimics the dynamics of a phase transition and also sets a general framework for the imposition of constraints on the $P_{\rm tr}$-$\Delta\mathcal{E}$ parameter space. It is worth commenting that the CSS parametrization is motivated by Nambu–Jona-Lasinio models which display nearly constant speed of sound (see Ref.~\cite{Alford-2013} and references therein). 

Finally, we adopt the APR~\cite{Akmal-1998,Baym-1971,Douchin-2001,Typel-2022} and the GRDF-DD2~\cite{Typel-2018} (simply DD2 from now on for practical purposes) EOSs for the description of the hadronic phase. It is worth noting that the APR model provides a softer representation of nuclear matter compared to DD2, and hence predicts compact star configurations with lower radii. 

\section{Radial oscillations and stellar stability} 

\subsection{Radial oscillation equations}
In a pioneering work, Chandrasekhar~\cite{Chandrasekhar-1964} was the first to extract the equations describing the radial oscillations of relativistic compact stars, with the aim of accessing their dynamical stability. Since then, a lot of different studies have been devoted to the investigation of radial perturbations, focusing on different theoretical assumptions for the nature of matter in the stellar interior~\cite{Glass-1983,Vath-1992,Gondek-1997,Kokkotas-2001,Gupta-2002,Flores-2010,Pereira-2018,Sen-2023,Rather-2024,Panotopoulos-2017}. 

In the present study we are going to follow the formalism developed by Gondek {\it et al.}~\cite{Gondek-1997} since it is quite suitable for numerical implementations. In this framework, the differential equations are formulated in terms of the relative radial displacement of the fluid $ \xi =\Delta r/r$ and the Lagrangian pressure perturbation $\Delta P$. Then, the set of ordinary differential equations (ODE) describing the radial perturbations is given, in geometrical units, as~\cite{Gondek-1997,Pereira-2018}

\begin{equation} \label{e2}
    \frac{d \xi}{dr} = V(r)\xi+W(r)\Delta P,
\end{equation}
\begin{equation} \label{e3}
    \frac{d\Delta P}{dr} = X(r) \xi+Y(r)\Delta P,
\end{equation}

where
\begin{equation} \label{e4}
    V(r) = -\frac{3}{r}-\frac{dP}{dr}\frac{1}{P+\mathcal{E}},
\end{equation}
\begin{equation} \label{e5}
    W(r) = -\frac{1}{r\Gamma P},
\end{equation}
\begin{equation} \label{e6}
\begin{split}
    X(r) & =\omega^2 e^{\lambda-\nu} (P+\mathcal{E})r-4\frac{dP}{dr} \\
    & +\left(\frac{dP}{dr}\right)^2\frac{r}{P+\mathcal{E}}-8\pi(P+\mathcal{E})Pre^{\lambda},
    \end{split}
\end{equation}

\begin{equation} \label{e7}
    Y(r) = \frac{dP}{dr}\frac{1}{P+\mathcal{E}}-4\pi(P+\mathcal{E})re^{\lambda}.
\end{equation}

In Eqs.~(\ref{e4})-(\ref{e7}), $P$, $\mathcal{E}$, $dP/dr$, $\lambda$ and $\nu$ derive from the solution of the Tolman-Oppenheimer-Volkov equations, which describe static and spherically symmetric relativistic stars.~Notably, $\lambda$ and $\nu$ are the functions found in the corresponding spacetime metric (for the unperturbed star), which is the following
\begin{equation}
    ds^2 = -e^{\nu(r)}dt^2+e^{\lambda(r)}dr^2+r^2(d\theta+\sin^2\theta d\phi^2).
\end{equation}
In addition, $\Gamma$ stands for the adiabatic index given by
\begin{equation}
    \Gamma=(1+\mathcal{E}/P) c_s^2,
\end{equation}
and $\omega$ is the oscillation mode frequency.

In order to solve the aforementioned ODE system, one needs an appropriate set of boundary conditions. The demand for regularity at the center of the star indicates that
\begin{equation}
    (\Delta P)_{r=0} = -(3\xi \Gamma P)_{r=0},
\end{equation}
while one can normalize the functions $\xi$ and $\Delta P$ so that $\xi(r=0)=1$. In addition, it is known that the pressure should vanish at all times as the radial distance reaches the radius of a star. As a consequence, the Lagrangian pressure perturbation should be zero at the stellar surface, and hence the second boundary condition is determined via
\begin{equation} \label{e11}
    (\Delta P)_{r=R} = 0.
\end{equation}

The system of Eqs.~(\ref{e2}) and (\ref{e3}) is solved through the implementation of the {\it shooting method}. In particular, for two trial values of the frequency $\omega$, we integrate the ODE system, from the center to the surface of the star, in order to obtain two values of the opposite sign for the Lagrangian pressure perturbation at the surface. Then, we employ a bracketing root finding algorithm to extract the frequency value for which the condition of Eq.~(\ref{e11}) is satisfied at the desired precision.~Notably, the aforementioned frequency is identified as an eigenvalue of the system (eigenfrequency), while the resulting functions $\xi$ and $\Delta P$ are identified as the corresponding eigenfunctions. By repeating this process, for different trial values, one obtains the frequencies of different radial oscillation modes, which can be ordered as
\begin{equation}
    \omega_0^2<\omega_1^2<...<\omega_n^2,
\end{equation}
where $n$ corresponds to the number of nodes appearing in the corresponding eigenfunctions. Interestingly, the stability of a star is determined by the sign of the derived $\omega^2$ values. In particular, a star is only stable if all of the eigenfrequencies are real numbers (i.e., if all of the resulting $\omega^2$ values are positive). As a result, if the {\it fundamental} mode (nodeless eigenfunctions) is characterized by a positive $\omega^2$ then stellar stability is ensured.
\\

\subsection{ Junction conditions at the hadron-quark interface.}\begin{figure*}[t]
  \centering  \includegraphics[width=\linewidth,scale=0.5]{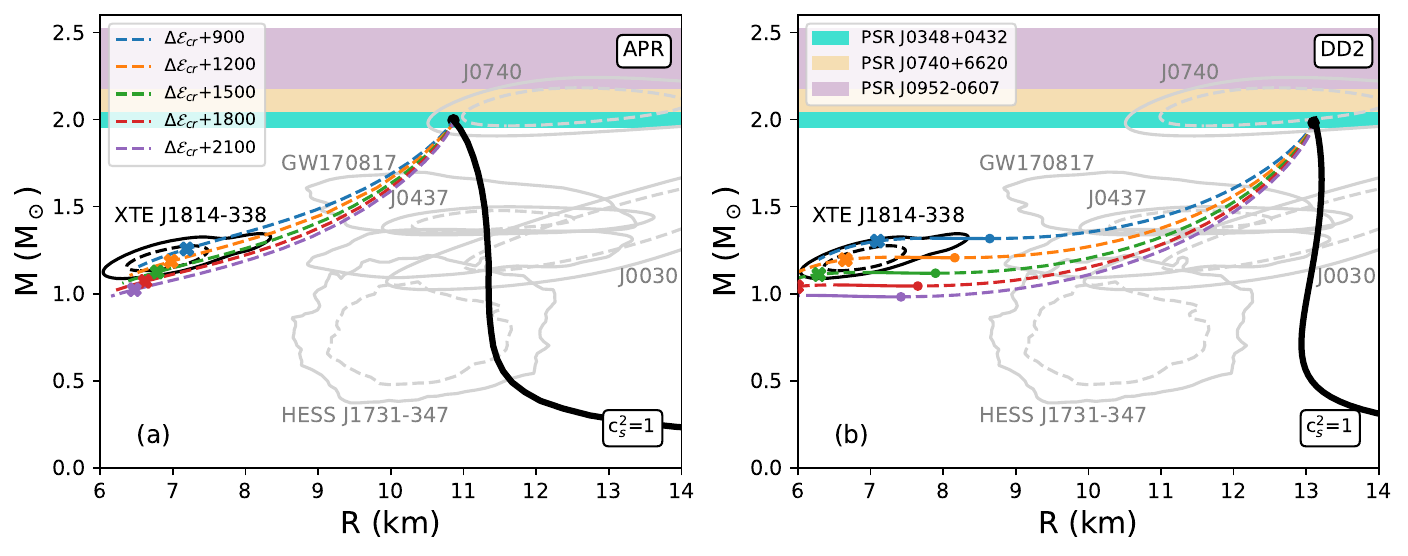}
  \caption{Mass-radius diagram for hybrid EOSs with different energy density jump values. Panel (a) contains the results for the case where the low density phase is described via the APR EOS, while for panel (b) the DD2 EOS was used. The phase transition takes place when the hadronic branch reaches the $2M_\odot$. The speed of sound is set equal to the speed of light ($c_s^2=1$). The solid parts of the curves indicate the regions where $dM/dP_c<0$, while for the dashed parts $dM/dP_c>0$. The colored circular points indicate the transition from a $dM/dP_c>0$ to a $dM/dP_c<0$ region. The x appearing on each curve denotes the last slow stable configuration.
  The gray contour regions denote Mass-radius measurements from PSR J0437–4715~\cite{Choudhury-2024}, PSR J0740+6620~\cite{Salmi-2024}, PSR J0030+0451~\cite{Vinciguerra-2024}, and
GW170817~\cite{Abbott-2017,Abbott-2018}. For PSR J0030+0451 we show the ST+PST NICER-only analysis~\cite{Vinciguerra-2024}, which results in similar estimations with the initial works of Refs.~\cite{Riley-2019,Miller-2019}.The black contour region indicates the mass and radius of the XTE J1814-338 pulsar~\cite{Kini-2024}. The confidence level that corresponds to the dashed (solid) contours is 68$\%$ (95$\%$). The shaded regions correspond to possible constraints on the maximum mass from the observation of PSR J0348+0432~\cite{Antoniadis-2013}, PSR J0740+6620~\cite{Cromatie-2020}, and PSR J0952-0607~\cite{Romani-2022}. The units of energy density jump are given in MeV fm$^{-3}$ which are omitted in the legend for simplicity.}
  \label{MR1}
\end{figure*}
In the case of hybrid stars, the fluid oscillation close to the region of the hadron-quark interface may lead to phase conversions~\cite{Pereira-2018}. The latter solely depends on the value of the phase conversion timescale $\tau_{\rm conv}$ compared to the perturbed fluid's oscillation period $\tau_{\rm osc}$. If $\tau_{\rm conv}$ is much larger than $\tau_{\rm osc}$, then the phase-splitting surface in the stellar interior oscillates with the same period as the perturbations. In contrast, it remains stationary when the oscillation period is larger than the conversion timescale. The first case is known in the literature as slow phase conversion, while the latter scenario is found as fast. 

As our main goal is to examine the stability of hybrid compact stars, where sharp discontinuities may be present, we need to use appropriate junction conditions when integrating Eqs.~(\ref{e2}) and (\ref{e3}). Depending on the considered assumption concerning the aforementioned timescales one has to use the following relations to accurately determine the oscillation mode frequencies of a hybrid compact star~\cite{Pereira-2018}:
\begin{enumerate}
    \item {\it Slow conversion}: 
    \begin{equation}
    \begin{split}
        \left[\xi\right]^+_-&=\xi^+-\xi^-=0 \hspace{0.3 cm} \mathrm{and} \\
        &\hspace{0.3 cm} \left[\Delta P\right]^+_-=\Delta P^+ - \Delta  P^-=0.
    \end{split}
    \end{equation}
    \item {\it Fast conversion}: \begin{equation}
        \left[\xi-\frac{\Delta P }{r P'}\right]^+_-=0 \hspace{0.3 cm} \mathrm{and} \hspace{0.3 cm} \left[\Delta P\right]^+_-=0,
    \end{equation}
\end{enumerate}
where $P'$ corresponds to the derivative of pressure with respect to the radial distance and the $+,-$ signs indicate the function values at each side of the phase-splitting surface.

The most usual way to study the stability of a stellar configuration is by examining the slope of the $M(P_c)$ curve (where $M$ stands for the stellar mass and $P_c$ for the central pressure). This method is known as the turning point criterion. In uniform stars a negative slope is always associated with unstable configurations. However, this is not always the case for hybrid stars (in the slow conversion scenario). In particular, it has been shown numerous times~\cite{Pereira-2018,Rather-2024,Lugones-2023,Pereira-2021,Flores-2012,DiClemente-2020,Rau-2023,Mariani-2022} that even for $dM/dP_c<0$ the frequency of the {\it fundamental} oscillation mode may be positive. As a consequence, hybrid configurations may be stable even at a descending branch of the $M-P_c$ curve. Notably, such configurations are often called $slow$ stable hybrid stars.

 \begin{figure*}[t]
  \centering  \includegraphics[width=\linewidth,scale=0.5]{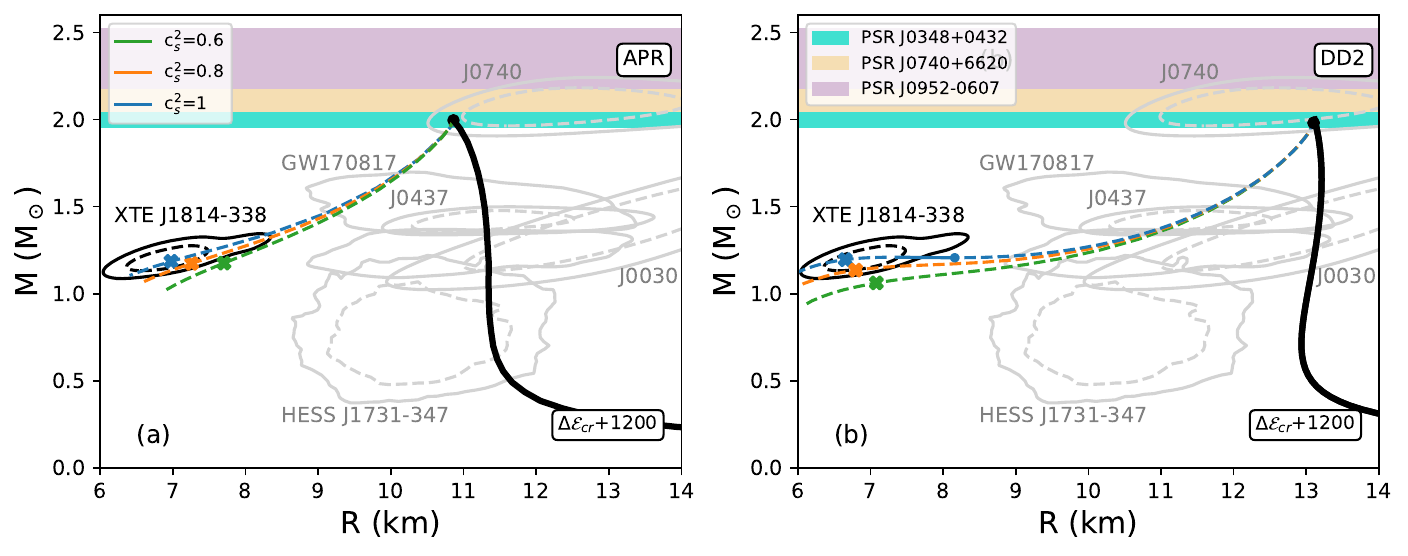}
  \caption{Mass-radius diagram for hybrid EOSs with different $c_s^2$ values. Panel (a) contains the results for the case where the low density phase is described via the APR EOS, while for panel (b) the DD2 EOS was used. The phase transition takes place when the hadronic branch reaches the $2M_\odot$. The energy density jump is equal to $\Delta\mathcal{E}_\mathrm{cr}+1200$ MeV fm$^{-3}$.The speed of sound is set equal to the speed of light ($c_s^2=1$). The solid parts of the curves indicate the regions where $dM/dP_c<0$, while for the dashed parts $dM/dP_c>0$. The colored circular points indicate the transition from a $dM/dP_c>0$ to a $dM/dP_c<0$ region. The x appearing on each curve denotes the last slow stable configuration. The observational constraints are the same as in Fig.~\ref{MR1}.}
  \label{MR2}
\end{figure*}
\begin{figure*}[t]
  \centering  \includegraphics[width=\linewidth,scale=0.5]{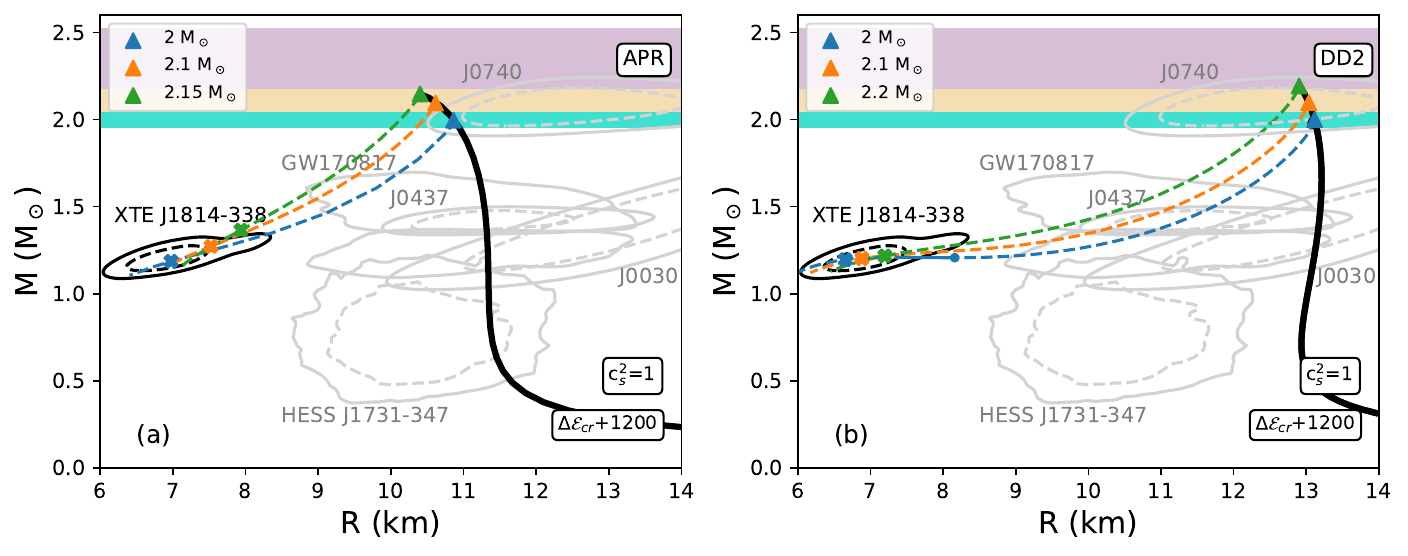}
  \caption{Mass-radius diagram for hybrid EOSs with different values of transition density (or different transition masses). Panel (a) contains the results for the case where the low density phase is described via the APR EOS, while for panel (b) the DD2 EOS was used. The speed of sound is set equal to the speed of light ($c_s^2=1$). The energy density jump is equal to $\Delta\mathcal{E}_\mathrm{cr}+1200$ MeV fm$^{-3}$. The speed of sound is set equal to the speed of light ($c_s^2=1$). The solid parts of the curves indicate the regions where $dM/dP_c<0$, while for the dashed parts $dM/dP_c>0$. The colored circular points indicate the transition from a $dM/dP_c>0$ to a $dM/dP_c<0$ region. The x appearing on each curve denotes the last slow stable configuration. The observational constraints are the same as in Fig.~\ref{MR1}}
  \label{MR3}
\end{figure*}
\begin{figure}[h!]
  \centering  \includegraphics[width=\linewidth,scale=0.5]{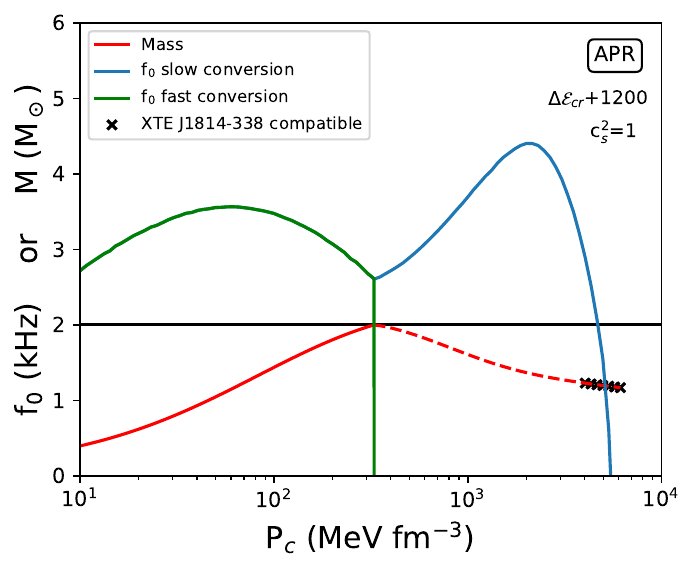}
  \caption{The frequency ($f=\omega/2\pi$) of the fundamental mode (green and blue curves) and the mass (red curve) as a function of the central pressure. The selected hybrid EOS of state derives through the combination of the APR model and the CSS parametrization. The speed of sound is equal to speed of light, while the energy density is equal to $\Delta\mathcal{E}_\mathrm{cr}+1200$ MeV fm$^{-3}$. The phase transition occurs right after the hadronic branch reaches the $2M_\odot$. The blue (green) curves stand for the case of slow (fast) phase conversion. The solid (dashed) parts of the red curve denote the region where the mass increases (decreases) with increasing central pressure.}
  \label{stability_apr}
\end{figure}
\begin{figure}[h!]
  \centering  \includegraphics[width=\linewidth,scale=0.5]{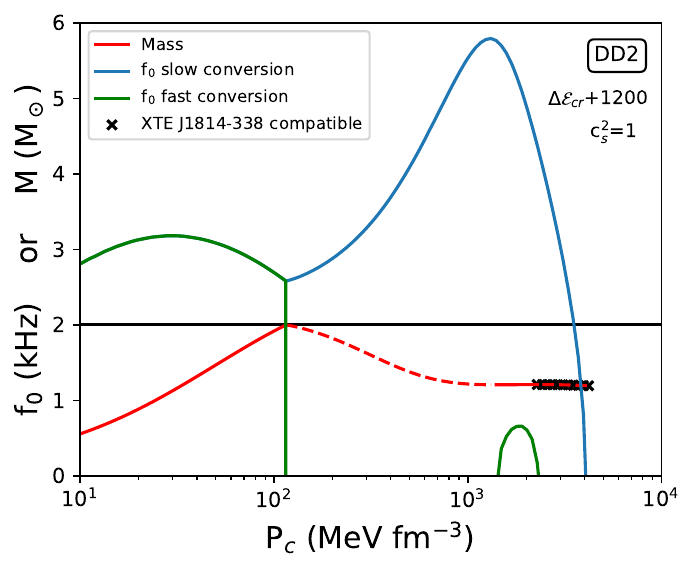}
  \caption{Same as Fig.~\ref{stability_apr} but for the DD2 EOS.}
  \label{stability_DD2}
\end{figure}
\section{Results and Discussion}
We aim to examine the explanation of the XTE J1814-338 properties by varying the quark EOS parameters. It is important to comment that the resulting hybrid EOSs ought to be consistent with NICER-inferred $M-R$ constraints from rotation-powered
millisecond pulsars. Notably, the employed hadronic models are both consistent with state-of-the-art constraints on the properties of PSR J0437–4715~\cite{Choudhury-2024}, PSR J0740+6620~\cite{Salmi-2024} and PSR J0030+0451~\cite{Vinciguerra-2024} (at least at the $2\sigma$ level). In addition, hybrid EOSs must be compatible with the observation of massive compact stars. For the latter reason, the phase transition is to occur beyond the value of pressure for which the hadronic EOS predicts stable stellar configurations with $M>2M_\odot$. Therefore, instead of describing an EOS in terms of the transition density, we will mark it by its transition mass ($M_\mathrm{tr}$, the mass of the heaviest purely hadronic configuration).

In Fig.~\ref{MR1} we plot the Mass-radius dependence predicted by different hybrid EOSs. In the derivation of the hybrid models we have set the speed of sound equal to the speed of light, while the phase transition takes place right after the hadronic branch (solid black curves) surpasses the two solar masses. In addition, we have varied the value of the energy density jump in the range [$\Delta\mathcal{E}_\mathrm{cr}+900,\Delta\mathcal{E}_\mathrm{cr}+2100$] MeV fm$^{-3}$, where $\Delta\mathcal{E}_\mathrm{cr}$ stands for the so-called critical energy density jump indicated by the work of Seidov~\cite{Seidov-1971,Alford-2013,Bielich-2020}.~This critical jump sets a benchmark for estimating the width of the density discontinuity needed for the appearance of a descending (hybrid) branch in the mass-central pressure diagram ($dM/dP_c<0$). Notably, the presence of these descending regions, which are depicted via the dashed parts of the curves, is crucial in our attempt of reproducing the XTE J1814-338 constraints. As expected, increasing the energy density jump leads to EOS softening and hence for a given stellar radius we obtain lower masses. 

One crucial point, regarding the effects of the selected hadronic EOS, is that for the DD2 model, which is stiffer, the descending hybrid branch is interrupted by a region where mass increases again with increasing central pressure (colored solid regions in Fig.~\ref{MR1}(b)).~As previously mentioned, the turning point criterion states that a configuration is only stable if it lies in a region of the $M-P_c$ diagram where the slope $dM/dP_c$ is positive.~Therefore, for the DD2 EOS, there are in fact stable configurations that are compatible with the existence of XTE J1814-338. However, in the scenario where the low density phase is modeled via a softer EOS, in our case APR, one can observe that all of the hybrid configurations that pass through the $1\sigma$ estimations on XTE J1814-338 are unstable when considering the turning point criterion. That is of particular importance, as the stiffer EOS is of course not compatible with the recent observation on the HESS J1731-347 remnant, while it is only marginally consistent with the GW170817 event. However, it is worth
mentioning that, the
HESS J1731-347 mass and radius estimation relies on certain assumptions~\cite{AlfordJ-2023}. In particular, it is thought that the star
has a uniform temperature carbon atmosphere, and it is
located at a distance of 2.5 kpc~\cite{AlfordJ-2023}.~Therefore, future cross checks that confirm the validity of the HESS J1731-347 constraints would be necessary to potentially rule out the DD2 EOS.

In Figs.~\ref{MR2} and~\ref{MR3} we wish to clarify the effects of varying the rest of the quark EOS parameters on the resulting hybrid configurations. More precisely, in Fig.~\ref{MR2} we have kept the transition density fixed (phase transition takes place when the hybrid branch reaches the two solar mass) and we have varied the value of the speed of sound. The energy density jump was also fixed at a value of $\Delta\mathcal{E}_\mathrm{cr}+1200$ MeV fm$^{-3}$, which was one of the cases that were found to be compatible with XTE J1814-338 in Fig.~\ref{MR1}. We observed that decreasing the speed of sound softens the EOS and therefore the explanation of XTE J1814-338 becomes more difficult. However, this problem could potentially be tackled by selecting a lower value for the energy density jump (which would lead to stiffening).~Another interesting point is that the induced softening does not allow for stable hybrid configurations (in terms of the turning point criterion) for both hadronic EOSs.

In Fig.~\ref{MR3}, we have kept the speed of sound and energy density jump fixed and we have varied the transition density. Once again, we observe that increasing the transition density eliminates the stable hybrid branch that appears for the DD2 EOS (when considering the turning point approach).

Up to this moment, we have only examined the stability of the resulting hybrid configurations based on the turning point criterion. However, as previously mentioned, the sign of the $dM/dP_c$ slope does not provide a sufficient stability condition when studying hybrid stars. In particular, in the so-called slow conversion scenario, hybrid configurations may be stable even at a descending branch of the $M(P_c)$ curve. For a recent view on the subject we refer to the work of Mariani {\it et al.}~\cite{Mariani-2024}, where the authors argued that the HESS J1731-347 remnant could potentially be a slow stable hybrid star. For the reasons above, we wanted to examine the possible stability of hybrid configurations, which are compatible with XTE J1814-338, by following the rigorous method of solving the radial oscillation equations. Thus, in Figs.~\ref{MR1}-\ref{MR3} one can find the last slow stable configurations (terminal configurations), for each of the hybrid models constructed in this work, marked with an x. Beyond these configurations (i.e., for larger central pressure) the eigenvalue of the {\it fundamental} mode becomes imaginary and therefore the derived stellar models are no longer stable.

\begin{table}[t]
\caption{\label{tab:table1}
The mass and radius of the last slow stable hybrid configuration (terminal configuration) for different transition masses (i.e., for different transition densities). The hadronic phase is described via the DD2 EOS, while $\Delta\mathcal{E}$ and $c_s^2$ are equal to $\Delta
\mathcal{E}_\mathrm{cr}+1200$ MeV fm$^{-3}$ and 1, respectively. The $M-R$ diagrams for the considered hybrid EOSs are shown in Fig.~\ref{MR3}(b).} 
\begin{ruledtabular}
\begin{tabular}{cccc}
EOS & $M_\mathrm{tr}$ ($M_\odot$)& $M^\mathrm{term}$ ($M_\odot$) & $R^\mathrm{term}$ (km)\\
\hline
DD2  & 2.0 & 1.197 & 6.68 
\\
     & 2.1 & 1.202 & 6.89 
\\
     & 2.2 & 1.218 & 7.20
\\
\end{tabular}
\end{ruledtabular}
\end{table}
In Figs.~\ref{stability_apr} and \ref{stability_DD2} we have plotted the frequency of the {\it fundamental} mode as a function of the central pressure. We have also included the $M-P_c$ curve which is denoted by red (the dashed regions are characterized by $dM/dP_c<0$). The blue curves depict the frequencies related to the~slow conversion scenario, while the green ones indicate the fast conversion case. With x, we have marked the configurations that are compatible with the XTE J1814-338 constraints. As is evident, both of the hadronic EOSs produce stable stellar configurations that can reproduce the XTE J1814-338 properties. In the case of the stiffer DD2 EOS, those configurations are stable in both~slow and fast conversion scenarios. However, in the case of the APR EOS the resulting stellar models are only~slow stable. 

So far we have discussed the stability of XTE J1814-338 compatible configurations considering that the transition mass is at the $2M_\odot$.~However, there have been recent measurements that suggest the existence of even more massive compact stars.~For instance, Romani {\it et al.}~\cite{Romani-2022}, reported that PSR J0952-0602, one of the fastest rotating pulsars, has a mass of $2.35\pm0.17 M_\odot$. For the latter reason, we wish to examine if the hybrid configurations (appearing in Fig.~\ref{MR3}), which are compatible with XTE J1814-338, are also stable. In Table~\ref{tab:table1} we report the properties of the last slow stable configuration for different transition masses [marked with x in Fig.~\ref{MR3}(b)]. As is evident, the terminal configurations are within the predictions for the XTE J1814-338 properties and therefore the existence of such an object is potentially compatible with the observation of massive compact stars. Notably, the results in Table~\ref{tab:table1} were derived by considering the DD2 EOS for the description of the hadronic phase. We found that in the case where the low density phase is modeled via the APR EOS, the terminal configurations for $M_\mathrm{tr}=2.1,$ $2.15M_\odot$ appear before the XTE J1814-338 region ($1\sigma$) is reached. Nonetheless, for $M_\mathrm{tr}=2.1M_\odot$ there are stable configurations that are compatible with XTE J1814-338 at the $2\sigma$ level (see Fig.~\ref{MR3}(a)).

\section{Summary} In this work we have argued on the explanation of the recently announced XTE J1814-338 properties through the occurrence of an extremely strong phase transition in dense matter. In particular, we have constructed hybrid EOSs by considering interestingly large values for the energy density jump. For the description of the hadronic phase we have used the GRDF-DD2 and APR EOSs, while for quark matter we have utilized the widely employed constant speed of sound ansatz. 

 By analyzing the $M-R$ diagrams for different parametrizations of the quark phase, we were able to find hybrid configurations that satisfy the XTE J1814-338 constraints. When considering the GRDF-DD2 model (which is stiffer compared to APR), those stellar models were also stable in terms of the well-known turning point criterion. This was not the case for the APR EOS. However, by performing a complementary analysis, based on the radial oscillations of compact stars, we were able to show that hybrid configurations (compatible with XTE J1814-338) involving the APR model may be stable in a slow conversion scenario. This is of particular importance, as XTE J1814-338, in combination with future observations (that will constrain the stiffness of the nuclear EOS), may not only provide evidence of a strong QCD phase transition but also shed light on the phase conversion dynamics.

The precise determination of this object's mass and radius may, in any case, unlock new and diverse theoretical models for describing its structure. Interestingly, these models may substantially deviate from the conventional framework used for the description of  pulsars.
\\\\
\begin{acknowledgements}The authors would like to thank Professor K.D. Kokkotas for his useful insights and comments and Mr Y. Kini for providing the data for the contour regions for XTE J1814-338. The research work was supported by the Hellenic Foundation for Research and Innovation (HFRI) under the 5th Call for HFRI PhD Fellowships (Fellowship No.~19175). 
\end{acknowledgements}

\end{document}